\documentclass[apjl,twocolumn]{emulateapj_mod}
\usepackage{epsfig,apjfonts,mathptmx}

\def\gtsima{$\; \buildrel > \over \sim \;$}
\def\ltsima{$\; \buildrel < \over \sim \;$}
\def\prosima{$\; \buildrel \propto \over \sim \;$}
\def\gsim{\lower.5ex\hbox{\gtsima}}
\def\lsim{\lower.5ex\hbox{\ltsima}}
\def\simgt{\lower.5ex\hbox{\gtsima}}
\def\simlt{\lower.5ex\hbox{\ltsima}}
\def\simpr{\lower.5ex\hbox{\prosima}}

\def\h1{$h^{-1}$}
\def\eeq{\end{equation}}
\def\beq{\begin{equation}}
\def\24mu{24\,$\mu{\rm m}$}
\def\70mu{70\,$\mu{\rm m}$}
\def\8mu{8\,$\mu{\rm m}$}

\shorttitle{A CO redshift for GN10}

\shortauthors{E. Daddi et al.}

\begin{document}

\title{
A CO emission line from the optical and near-IR undetected submillimeter galaxy GN10
}

   \author{E. Daddi\altaffilmark{1},
           H. Dannerbauer\altaffilmark{2},
           M. Krips\altaffilmark{3},
	   F. Walter\altaffilmark{2}
           M. Dickinson\altaffilmark{4},
           D. Elbaz\altaffilmark{1},
           G.E. Morrison\altaffilmark{5,6}
           }

\altaffiltext{1}{
    CEA, Laboratoire AIM, 
    Irfu/SAp, F-91191 Gif-sur-Yvette, France
    [e-mail: {\em edaddi@cea.fr}]}
\altaffiltext{2}{MPIA, K\"onigstuhl 17, D-69117 Heidelberg, Germany}
\altaffiltext{3}{IRAM, St. Martin d'H\`eres, France}
\altaffiltext{4}{NOAO,  950 N. Cherry Ave., Tucson, AZ, 85719}
\altaffiltext{5}{IfA, University of Hawaii, Honolulu, HI 96822}
\altaffiltext{6}{CFHT, Kamuela, HI 96743}

\begin{abstract}

We report the detection of a CO emission line from the submillimiter galaxy (SMG) GN10 in the GOODS-N
field.
GN10 lacks any counterpart in extremely deep optical and
near-IR imaging obtained with the Hubble Space Telescope
and ground-based facilities. This is a
prototypical case of a source that is extremely obscured by dust, 
for which it is
practically impossible to derive a
spectroscopic redshift in the optical/near-IR.
Under the hypothesis       
that GN10 is part of a proto-cluster structure previously
identified at $z\sim4.05$ in the same field, we 
searched for CO[4-3] at 91.4~GHz with the IRAM Plateau de Bure Interferometer, and successfully
detected a line.
We find that the
most likely redshift identification is $z=4.0424\pm0.0013$, based on: 1) the very low chance
that the CO line is actually serendipitous from a different redshift; 2) a radio-IR photometric
redshift analysis; 3) the identical radio-IR SED, within a scaling factor, of two other SMGs
at the same redshift. 
The faintness at
optical/near-IR wavelengths requires an 
attenuation of
$A_{\rm V}\sim5$--7.5 mag. 
This result  supports the case 
that a substantial population of very high-$z$ SMGs exists
that had been missed by previous spectroscopic surveys. This is the
first time that a CO emission line has been detected for a galaxy that is invisible in the optical and near-IR.
Our work demonstrates the power of
existing and planned facilities for completing the census of star
formation and stellar mass in the distant Universe by measuring redshifts of the most
obscured galaxies through millimeter spectroscopy.

\end{abstract}


\section{Introduction}

Dust extinction at UV and optical rest frame wavelengths is a major
obstacle for obtaining a complete sampling of star formation in the
distant Universe.
GN10 (also known as GOODS 850-5) is one of the most striking
examples of an extremely obscured dusty
galaxy. Discovered as a submillimeter emitting galaxy (SMG) by Wang,
Cowie and Barger (2004; $S(850\mu m)=12.9\pm2.1$ mJy), it was
later confirmed as one of the brightest galaxy in the GOODS-N region
at wavelengths between $850\mu$m and 1.25mm (Pope et al.\ 2006;
Dannerbauer et al.\  2008; Greve et al.\ 2008; Perera et al.\ 2008). Its
accurate position on the sky was obtained through interferometric observations of the dust continuum by
Wang et al.\ (2007)
with SMA at 870$\mu$m and by 
Dannerbauer et
al.\ (2008) at 1.25mm with the IRAM Plateau de Bure Interferometer (PdBI)
as well as in the radio at 1.4~GHz with the Very Large Array. 
The dust continuum emission in
the submillimeter and millimeter regime corresponds to an extreme luminosity of order
$10^{13}L_\odot$ and a star formation rate of order 1000
$M_\odot$~yr$^{-1}$, provided that it is at $z>1$. 
Despite
its huge ongoing star formation activity, the galaxy is undetected in
the highly sensitive Hubble Space Telescope (HST)  ACS imaging of Giavalisco et al.\ (2004) down
to limits of AB$\sim29$ magnitudes, and the galaxy
is also undetected to similar levels in the deep near-IR imaging with
Subaru and HST+NICMOS of Wang et al.\ (2009), while a faint but 
significant emission is seen in the GOODS Spitzer+IRAC data (Pope et al. 2006;
Dickinson et al.\, in preparation).

Dannerbauer et al.\ (2008) estimated a redshift of $z\sim4$ for this galaxy 
using spectral energy distribution (SED) fitting and employing the radio-IR relation,
while Wang et al.\ (2007) suggested an even higher $z\sim6$, interpreting the
very red $K - 3.6\mu$m color as being due to a strong, redshifted Balmer/4000\AA\ break
from an evolved stellar population.
Due to its faintness, it is practically impossible to measure the redshift of GN10 by
ordinary means of optical or near-IR spectroscopy. 
A MIPS flux density of $S(24\mu m)=30\mu$Jy makes it 
impossible to obtain a redshift with the Spitzer Infrared Spectrograph (IRS), 
a technique that has been successfully used for SMGs
(Men{\'e}ndez-Delmestre et al.\ 2007; Pope et
al.\ 2008).
The search for CO emission lines, successfully detected
in SMGs (review by 
Solomon \& van den Bout 2005), seems to be the only way of
measuring a redshift for GN10 before the launch of JWST or
the realization of $>30$m optical/near-IR telescopes. 
However, the most sensitive radio and millimeter
facilities have quite limited bandwidths, which complicate blind line
searches. For example, 3mm
observations with the PdBI cover the range of $\Delta
z/(1+z)\sim0.01$, requiring an accurate foreknowledge of the redshift
for CO observations.

\begin{figure}
\centering
\includegraphics[width=7.0cm,angle=-90]{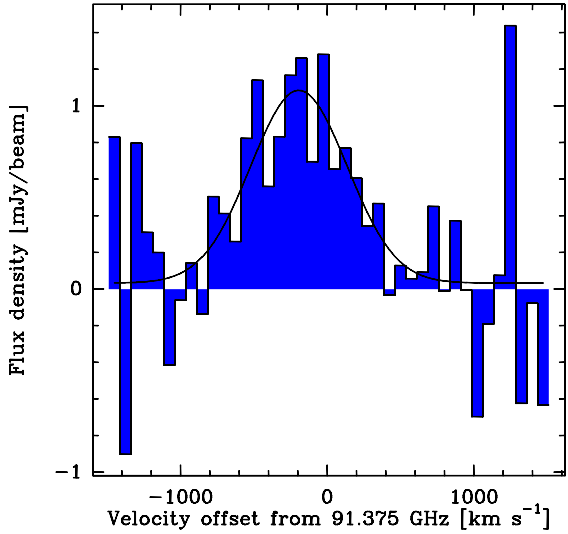}
\caption{Spectrum of GN10
sampled in steps of 75~km~s$^{-1}$. A Gaussian fit shows the 
presence of an emission line centered at $-192\pm70$~km~s$^{-1}$, corresponding to 
$z=4.0424\pm0.0013$ for CO[4-3], and no significant amount of continuum emission.
}
\label{fig:1}
\end{figure}

Daddi et al.\ (2009; D09 henceforth) proposed a new radio-IR
photometric redshift technique capable of obtaining accuracies of
$\Delta z/(1+z)\sim0.1$ for SMGs. 
However, this accuracy is still not sufficient for blind CO follow-up.
D09 also reported the discovery of a proto-cluster
structure at $z=4.05$ in GOODS-N that includes two SMGs (GN20
and GN20.2a).  GN10 has a radio-IR photometric redshift consistent
with being part of this proto-cluster structure.

In this letter we report on IRAM PdBI observations of GN10, searching
for a possible emission line of CO at $z\sim4.05$ based on the
hypothesis that GN10 is indeed part of the proto-cluster structure.

\section{Observations, data reduction and analysis}

We observed GN10 with the IRAM PdBI using the full array (with 6
antennas) for 4.0 hours on source on May 27th 2008 in D-configuration
(synthesized beam of $6.6''\times5.4''$), for 3.8 hours on November
27th 2008 in the C-configuration (synthesized beam of
$5.4''\times3.0''$), and for 6.8 hours on December 29th 2008 and
January 5th 2009 in B-configuration (synthesized beam of
$1.7''\times1.3''$).  A tuning frequency of 91.375~GHz was used, with
a correlator setup yielding a bandpass of 1~GHz with two
polarizations. In the B- and D-configuration observations GN10 was 
19.7$''$ away from phase center, suffering a primary beam attenuation
(PBA) of 30\%. Those data were obtained while observing a galaxy
at $z=1.52$ as part of a CO survey of normal galaxies described in 
Daddi et al.\ (2008; and in preparation). 
In the C-configuration observations the
phase center was at 9.7$''$ away from GN10, leading to a 
PBA of 8\%.

\begin{figure}
\centering
\includegraphics[width=8.0cm,angle=0]{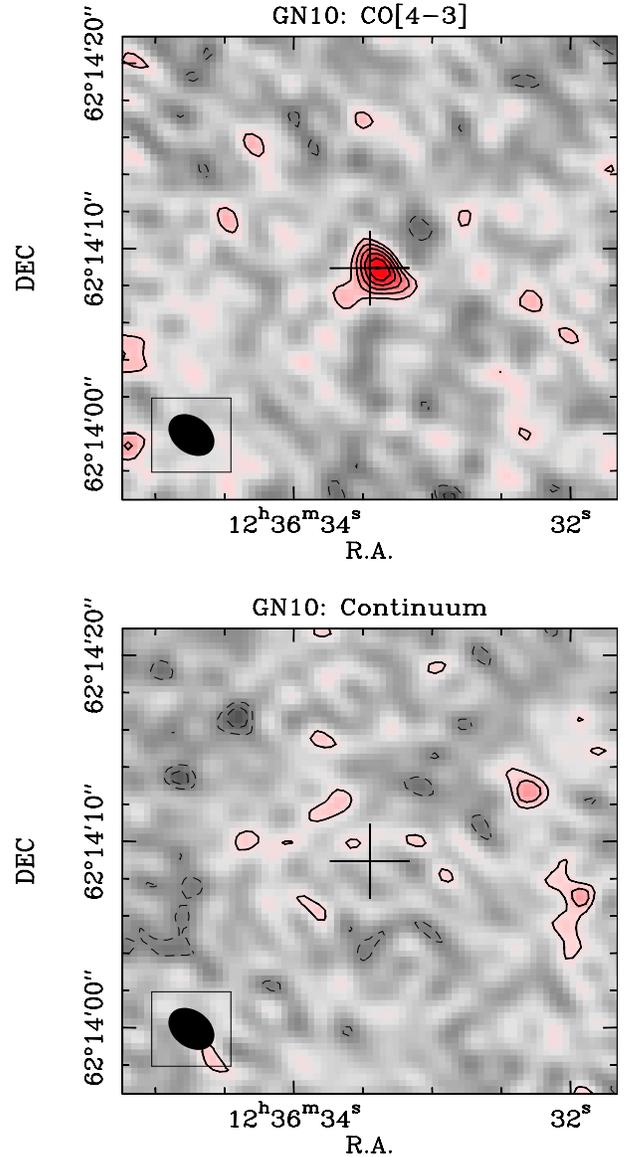}
\caption{Maps of the IRAM PdBI observations of GN10.
All available data from the B-, C- and
D-configurations were combined and natural weighting was used resulting in a beam
of $2.6''\times1.9''$ (insert in the bottom left corners). 
The 1.25mm dust continuum position from
Dannerbauer et al.\ (2008) 
is shown (cross with $\pm2''$ scale). 
The top panel shows the cleaned map of the CO emission averaged over a velocity range of -812.5~km~s$^{-1}$ to 
387.5~km~s$^{-1}$ (zero velocity corresponds to 91.375~GHz). The bottom panel shows the average of the remaining
velocity range.
Red is for positive and gray is for negative signal. Contours are shown starting 
at $\pm2\sigma$
in steps of $1\sigma$ (96$\mu$Jy beam$^{-1}$ and 79$\mu$Jy beam$^{-1}$ for line and continuum maps, respectively).
}
\label{fig:2}
\end{figure}
We reduced the data with the GILDAS software packages CLIC and MAP,
similarly to what is described in D09 and Daddi et al.\ (2008). The
maps obtained using natural weights
have noise levels (for the full 1~GHz bandpass) of 113$\mu$Jy~beam$^{-1}$,
87$\mu$Jy~beam$^{-1}$ and 63$\mu$Jy~beam$^{-1}$ for the D-, C- and
B-configuration data, respectively.

We extracted spectra by independently fitting the reduced uv data of
the D-, C- and B-configuration observations sampled with
75~km~s$^{-1}$ spectral bins, using a point source model.  We
corrected each spectrum for the PBA and averaged them with weighting
according to the different noise levels.  Fig.~\ref{fig:1} shows the
result: significant emission is detected, consistent with a CO
emission line centered at $-192\pm70$~km~s$^{-1}$ (all velocities are
given relative to the tuning frequency) and with a
velocity FWHM of $770\pm200$~km~s$^{-1}$. Integrating the spectrum
from -812.5~km~s$^{-1}$ to 387.5~km~s$^{-1}$ (i.e., a total 
bandwidth of 1200~km~s$^{-1}$)
results in a detection with $S/N>7$ and a flux density of
$0.74\pm0.10$~mJy. Integrating the spectrum outside the detected
emission line results in a continuum estimate of $0.02\pm0.09$~mJy.
Assuming this value for the continuum emission, and accounting for its 
uncertainty, we
derive an integrated flux of $0.86\pm0.16$~Jy~km~s$^{-1}$ for the
emission line, which is consistent with what found from the Gaussian
fit shown in Fig.~\ref{fig:1}.

\begin{figure}
\centering
\includegraphics[width=8.0cm,angle=0]{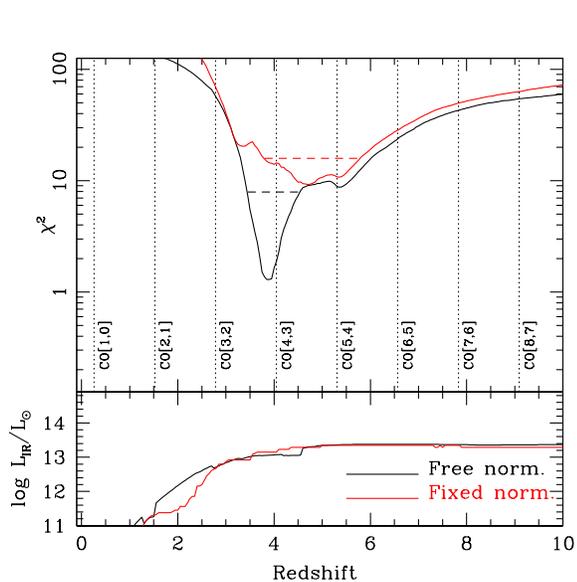}
\caption{Radio-IR photometric redshift of GN10, using the technique described
in D09. The top panel shows ${\chi}^2$ versus redshift and the bottom panel
the implied total IR luminosity. The black solid line shows the results allowing
for a renormalization of the CE01 models in the fitting,
while the red solid line
was obtained by preserving the intrinsic normalization of the CE01 models. The dashed
horizontal lines correspond to the 99\% confidence level ranges. The vertical dotted
lines show the permitted redshifts, 
given the detection of a CO line at 91.4~GHz.
}
\label{fig:3}
\end{figure}

Fig.~\ref{fig:2} shows the line (top) and continuum (bottom) maps
obtained combining the data from all configurations using
natural weighting (resulting in a synthesized beam of
$2.6''\times1.9''$).  The emission line is clearly detected, only
0.4$''$ to the West of the 1.25mm continuum position of GN10 measured
by Dannerbauer et al.\ (2008; see cross in Fig.~2,
top), a small offset which is not significant.

The line does not appear to be significantly resolved. A fit to the
combined B-, C- and D- configuration visibilities with a circular
Gaussian model results in a FWHM$=0.6''\pm0.3''$.

\begin{figure}
\centering
\includegraphics[width=8.0cm,angle=0]{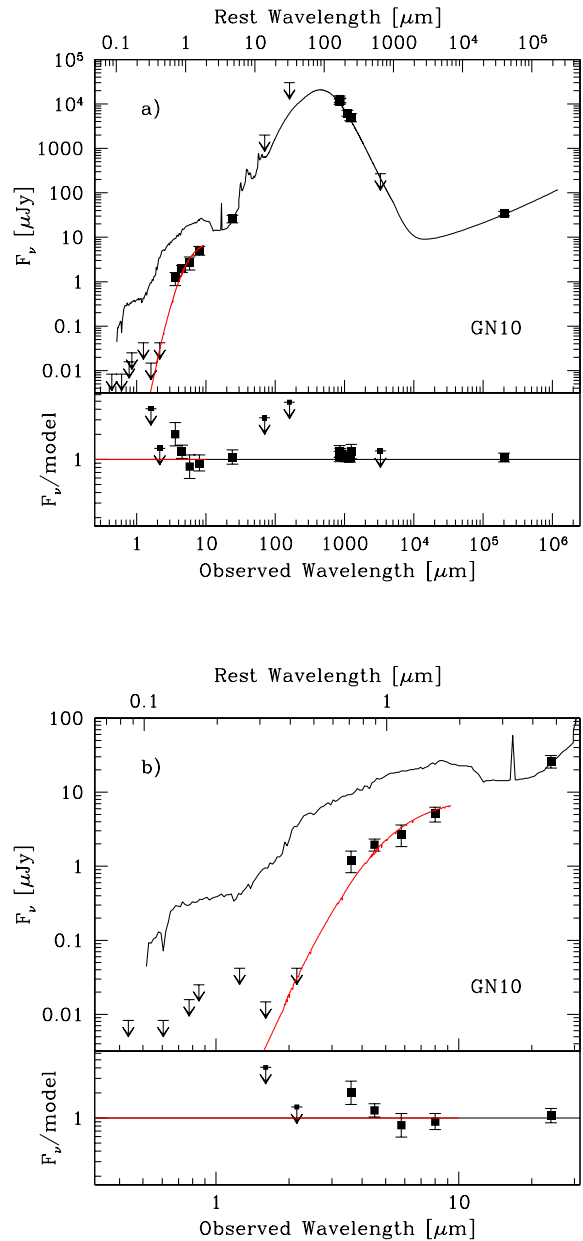}
\caption{a) SED of GN10 with all the available flux 
density measurements from the UV to radio (Wang et al.\ 2009; Dannerbauer et al.\ 2008; Perera et al.\ 2008;
Greve et al.\ 2008; this paper; all upper limits are shown at $3\sigma$). The 
solid line is a model from the CE01 library redshifted to $z=4.04$ and having an intrinsic 
$L_{\rm IR}=1.44\times10^{12}L_\odot$, scaled up by a factor of 8.3 to a
luminosity of $L_{\rm IR}=1.2\times10^{13}L_\odot$. This is the best--fitting
model for the dusty SED observed from 24$\mu$m to 1.4~GHz. The red solid line
is a star forming galaxy model from the Maraston (2005) library, reddened with
the Calzetti law and $A_V=7.5$. The lower panel shows the available
observations divided by the models (the CE01 model for wavelengths above 
10$\mu$m and the Maraston model below 10$\mu$m).
b) a zoom of the region below 10$\mu$m.
}
\label{fig:4}
\end{figure}
\section{Redshift identification}

The detection of our
single CO emission line at 91.4~GHz fixes the redshift as one of $1 + z_{CO}
\sim 1.261\times n$, in the case that the observed transition is
$CO[(n)-(n-1)]$
\footnote{
Detections of other lines (e.g., CI or HCN/HCO+) are unlikely, as they are typically much fainter than the CO
lines}.
We find that there are several convincing reasons supporting 
the identification of the line with CO[4-3], corresponding to $z=4.0424\pm0.0013$.
This is based on 3 lines of argument, discussed in detail in the following.

First, this line is {\em not} serendipitously detected.
We obtained these PdBI observations in order to verify if GN10
is part of the $z\sim4.05$ proto-cluster structure identified by D09
in the GOODS-N field.
Our observations searched for possible CO[4-3] emission over the fairly
narrow redshift range of $4.02<z<4.07$.
The $z=4.042$ would place GN10 well
inside the velocity range of the proto-cluster structure in GOODS-N
(see Fig.~13 in D09)  which also includes two other CO detected SMGs (GN20 at
$z_{\rm CO}=4.055\pm0.001$ and GN20.2a at $z_{\rm CO}=4.051\pm0.003$; D09), as well as 12
other optically-selected galaxies with 
concordant optical spectroscopic redshifts (D09; Stern et al., in preparation). 
The probability of a CO line being detected by chance in our 1~GHz bandwidth
is of order of 2\%. We thus consider it unlikely that we might have detected
a different CO transition from another redshift.

\begin{deluxetable*}{cccccccccc}
\tabletypesize{\scriptsize}
\tablecaption{Properties of GN10}
\tablewidth{0pt}
\tablehead{
\colhead{R.A.$_{\rm CO}$} &
\colhead{DEC$_{\rm CO}$} &
\colhead{z$_{\rm CO}$} &
\colhead{$I_{\rm CO}$} &
\colhead{$\Delta v_{\rm FWHM}$} &
\colhead{$L'_{CO[4-3]}$}&
\colhead{$L_{\rm IR}$}&
\colhead{$M_{stars}$}&
\colhead{$M_{gas}$}&
\colhead{$A_{\rm V}$}
\\
\colhead{J2000}&
\colhead{J2000}&
\colhead{}&
\colhead{Jy km~s$^{-1}$}&
\colhead{km~s$^{-1}$}&
\colhead{K~km~s$^{-1}$~pc$^2$}&
\colhead{$L_\odot$}&
\colhead{$M_\odot$}&
\colhead{$M_\odot$}&
\colhead{mag}
}
\startdata
12:36:33.389 & 62:14:08.94 & 4.0424$\pm$0.0013 & 0.86$\pm$0.16 & 770$\pm$200 & 3.4$\pm$0.6$\times10^{10}$ & $1.2^{+0.7}_{-0.5}\times10^{13}$ & $\approx10^{11}$ & 2.7$\pm$0.5$\times10^{10}$ & 5.0--7.5
\enddata
\tablecomments{
Coordinates are from the CO[4-3] emission. Their formal error  
is 0.1$''$ for both R.A. and DEC. All CO-related
quantities are based on the measurement of the CO[4-3] emission. The 
molecular gas mass derivation  assumes
  that the CO[1-0] and CO[4-3] transitions 
have the same brightness temperature and a conversion factor 
$\alpha_{\rm CO}=0.8$~$M_\odot$~(K~km~s$^{-1}$~pc$^2$)$^{-1}$.
The stellar mass is fairly uncertain (see text), given the extreme reddening of
the source. $A_{\rm V}$ is given for a Calzetti et al.\ (2000) reddening law.
}
\label{tab:GN10_est}
\end{deluxetable*}

The line identification is also supported by the radio-IR photometric redshift
technique of D09,
calibrated on $>40$ SMGs with $1<z<4.5$ (of which 3 at $4<z<4.5$)
 and found to provide
an accuracy of $\Delta z/(1+z)\sim0.1$. 
The 1.25mm measurement from Greve et al.\ (2008;
$S(1.2{\rm mm})=4.9\pm0.7$mJy) is consistent with those at similar
wavelengths from Dannerbauer et al.\ (2008; $S(1.25{\rm mm})=5\pm1$mJy) and
Perera et al.\  (2008; $S(1.1{\rm mm})=6.23\pm0.97$mJy), and is included in
our analysis as well as the new 3mm
upper limit ($S(3.3{\rm mm})<0.27$~mJy; $3\sigma$) obtained in this work.
We use these in addition to the $24\mu$m, 850$\mu$m and 1.4~GHz flux density measurements
that were already considered in D09.
We compared the radio-IR observations of GN10 to template SEDs of
local IR luminous galaxies taken from the Chary \& Elbaz (2001)
library (CE01 henceforth).  
Fig.~\ref{fig:3} shows the results. 
When the intrinsic normalization of the
CE01 templates is kept fixed, the 99\% confidence range for the
radio-IR photometric redshift is consistent with the identification
of the emission line as either CO[4-3] at $z=4.042$ or CO[5-4] at $z=5.3$. 
Treating the CE01 templates normalization as a free parameter, we can
obtain a much better fit ($\chi^2=1.2$ versus $\chi^2=9.8$).
In this case,  CO[4-3] at
z=4.042 is the only identification consistent with the photometric redshift.
Based on the expected accuracy of this radio-IR photometric redshift technique,
this analysis is not consistent with the higher redshift ($z\simgt6$) proposed by Wang et al.\ (2009),
but it is consistent with the estimate from Dannerbauer et al.\ (2008).
The best--fitting CE01 template has an intrinsic $L_{\rm
IR}=1.44\times10^{12}L_\odot$ (throughout the paper we assume a WMAP5
cosmology), renormalized by the fitted factor of 8.3 to a
luminosity of $L_{\rm IR}=1.2\times10^{13}L_\odot$.  The fixed-normalization 
approach yields a very similar value for the infrared luminosity, as can
be seen in the bottom panel of Fig.~\ref{fig:3}.  
It is reassuring that the same CE01 template was also found to 
provide the best fit for GN20, at nearly the same redshift (D09). 
This is consistent with the general result that
distant SMGs have spectral shapes (or color ratios) in the radio-IR that are similar to
those of $z<0.5$ galaxies with $L_{\rm IR}\sim10^{12}L_\odot$.  
Fig.~\ref{fig:4} shows the best--fitting CE01 template overlaid on 
the observed photometric measurements for GN10. The bottom
panel shows the ratio between the model and the observed data. For
wavelengths larger than 10$\mu$m, the dusty CE01 template provides a
good fit to the data.  For wavelengths shorter than $10\mu$m,
the measured flux densities of the galaxy are fainter than the model
prediction, indicating that the source is very heavily obscured
by dust, as discussed in the next Section.

\noindent
Finally, we note that the radio-IR  SED of GN10 is virtualy identical to those of GN20 and GN20.2a
that are securely confirmed to be at $z=4.05$ (D09).
Scaling them down by a factor of 1.8, all the flux measurements for GN20
at radio-IR wavelengths agree with GN10 within the
uncertainties, including the CO[4-3] emission line flux. 
A similar
picture holds for GN20.2a (the exception is its
1.4~GHz flux that is likely affected by AGN emission, see D09).
This is a remarkable degree of uniformity, to the
$\sim20$\% level of the various measurement errors, in the dust
continuum and CO properties. Given that the radio-IR flux ratios are in general a very strong
function of redshift, this is strong evidence that the redshift of GN10 is very close to
that of GN20 and GN20.2a.

Although a detection of additional emission lines of CO will be required to definitely
confirm the redshift identification, 
we conclude that, based on the 3 arguments presented above, there is already compelling evidence for identification
of the line as CO[4-3] at $z=4.042$.  We assume this identification in the remainder of the paper.

\section{Discussion}

Identification of the line as CO[4-3] at $z=4.042$ implies a CO
luminosity of $L'_{\rm CO}=3.4\times10^{10}$~K~km~s$^{-1}$~pc$^2$.  The
IR to CO luminosity ratio of GN10 is $L_{\rm IR}/L'_{\rm CO[4-3]}=350^{+300}_{-160}$~$L_\odot$/(K~km~s$^{-1}$~pc$^2$). 
This is very similar to those
of GN20 and GN20.2a (D09). Using a conversion factor of $\alpha_{\rm
CO}=0.8$~$M_\odot$~(K~km~s$^{-1}$~pc$^2$)$^{-1}$, 
we estimate a total molecular gas mass of $2.7\times10^{10}M_\odot$ 
assuming that CO[4-3] is thermalized.

The radio flux of $34.4\pm4.2 \mu$Jy (Dannerbauer et al.\ 2008)
corresponds to $L_{\rm 1.4 GHz} =
3.8\times10^{24}$~W~Hz$^{-1}$. Together with our estimate of $L_{\rm
IR}=1.2\times10^{13}L_\odot$, GN10  lies on
the local radio-IR correlation (e.g., Yun et al.\ 2001), as found also for GN20 (D09).

While the SEDs of GN10, GN20 and GN20.2a are very similar at radio-IR wavelengths, 
the galaxies are quite different at 
wavelengths shorter than 10$\mu$m, where the light is
dominated by the emission of stars. 
GN20 and GN20.2a are classified as B-band dropouts in the HST+ACS
imaging, and both
are relatively luminous and blue in the UV rest frame.
This indicates a relatively low extinction by dust.
Instead, GN10 is undetected at UV and optical rest
frame wavelengths, implying overall a very strong extinction of the
stellar light by dust.

It seems likely that the dust extinction in  these galaxies
is not homogeneous. Among the few known SMGs with confirmed
CO redshifts $z>4$, GN20, GN20.2a and the $z=4.54$ SMG by 
Capak et al.\ (2008; see also Schinnerer et al.\ 2008) all have significant
offsets of order of $0.5''$ or more between the detected (and
relatively blue) UV emission and the radio or CO positions. This
suggests a complex    dust distribution with `holes' that, in some cases, 
allow us to directly observe the emission from stars. In the case of GN10, 
however, the UV/optical light seems to be wholly obscured along our line of sight. 

We now discuss what levels of extinction and reddening are required to
explain the observed properties of GN10.  We can use the photometry
comparison to GN20 and GN20.2a to derive a first guess of the amount
of extinction present, assuming that to first order 
the stellar mass to IR continuum luminosity ratio is the same
within these galaxies. Normalizing the galaxies SEDs at wavelengths
longer than 10$\mu$m, GN10 is found to be $1\pm0.2$ mag fainter in
the IRAC bands (observed 0.7--1.6$\mu$m rest frame) than both GN20 and
GN20.2a. For a Calzetti et al.\ (2000) attenuation law this corresponds
to $A_{\rm V}\sim5$ mag. If we fit the UV rest frame to near-IR
rest-frame SED with constant star formation models from the Maraston
(2005) library, this requires an even larger $A_{\rm V}\sim7.5$ mag,
mainly to be consistent with the very deep flux limit from the K-band
observations of Wang et al.\ (2009).  

Fig.~\ref{fig:4} shows this fit (red curve in the top panel) and its
residuals (bottom panel).  
The fit is redder than the trend suggested by the IRAC
bands. 
Fitting only the IRAC bands would suggest again a reddening of
$A_{\rm V}\sim5$ mag. The K-band flux density upper
limit by Wang et al.\ (2009) shown in Fig.~\ref{fig:4} is derived for a 0.6$''$ aperture, but the
stellar light might be well distributed over a larger region.  In such
a case a less stringent K-band upper limit would be allowed, and
solutions with less attenuation than $A_{\rm V}\sim7.5$ mag would
become plausible.  We conclude that there is evidence for extreme
obscuration, likely at some level between $A_{\rm V}\sim5$--7.5 for a
Calzetti law. The fits imply a stellar mass of about
$1\pm0.5\times10^{11}M_\odot$ (Chabrier IMF) for GN10. Given the
case for extreme obscuration, this stellar mass estimate
should be treated with some caution. 

We note that the large CO[4-3] line width of about 770~km~s$^{-1}$ (very
similar to those for GN20 and GN20.2a) suggests a large dynamical mass,
assuming that the size of the CO emission is typical (e.g., a few kpc; Tacconi et
al.\ 2006). This would be consistent with the stellar mass that we
derive.  That stellar mass for GN10 is 3 times larger than the nominal
completeness limit for the IRAC-selected sample of massive galaxy 
candidates at $z>3.5$ from Mancini et al.\ (2009). However, GN10 is
absent from the Mancini sample because it is fainter than the adopted 
IRAC magnitude limit ($m_{4.5\mu m}<23$ AB), due to its extreme 
attenuation.  This illustrates how dust obscuration can be a significant 
limitation for constructing complete samples of galaxies in the distant 
Universe, even when using surveys based on Spitzer IRAC data.

On the basis of the identitication of the line as CO[4-3] at $z=4.042$,
GN10 is now the 
third SMG in the
$z=4.05$ GOODS-N proto-cluster structure. Our finding also supports the suggestion
that there is indeed a substantial population of SMGs $z>4$ (e.g., Dannerbauer et al.\ 2002; 
2004; 2008; Dunlop et al.\ 2004;
Wang et al.\ 2007; 2009; Younger et al.\ 2007; Capak et al.\ 2008; Coppin et al.\ 2009; D09),
with several others awaiting to be spectroscopically confirmed
in GOODS-N (D09).  Contrarily to GN20 and GN20.2a, GN10 is not
surrounded by an excess of B-band dropout Lyman break galaxies: only
two such objects are present within 25$''$ from GN10, while 14 are
found within the same distance from GN20.  Of these two B-band dropout  galaxies one
has an optical spectroscopic redshift of $z=4.053$ (Stern et al., in preparation; based on the
detection of Ly$\alpha$ emission with keck+DEIMOS),
placing it in the proto-cluster
structure, while the other has no known spectroscopic redshift.
The on-sky separation between GN10 and
GN20 is 9$'$, or 4.0~Mpc comoving. The velocity separation of $\Delta
v=3900\pm600$~km~s$^{-1}$ (from $\Delta z=0.013\pm0.002$) corresponds
to a line-of-sight distance of $9.3\pm1.2$~Mpc comoving (this is likely an upper limit, 
as part of the velocity separation
could be due to peculiar velocities within the structure). GN10 is closer
to GN20.2a both in the sky and in velocity space.  It appears that the
proto-cluster structure is fairly extended on the sky, and presumably
several other vigorous starburst galaxies
could also be part of this structure, as also suggested by D09. More   
work and observations are required for a full characterization of this
interesting high-$z$ overdensity of galaxies.

This is the first time that the redshift of a source that is undetected at
optical or near-infrared wavelengths has been derived through
measurement of CO lines. Our result clearly demonstrates that this will be 
a very powerful technique to identify the earliest and most obscured
star forming galaxies, in particular once wider band and/or more sensitive 
instruments will be available, such as the upcoming 8~GHz receiver upgrade
for PdBI, the Redshift Receiver for the Large Millimeter Telescope (LMT), 
and the Atacama Large Millimeter Array (ALMA).

\acknowledgements 
Based on observations with the IRAM
Plateau de Bure Interferometer. IRAM is supported by INSU/CNRS
(France), MPG (Germany) and IGN (Spain). 
We acknowledge funding
ANR-07-BLAN-0228 and ANR-08-JCJC-0008 and
NASA
support, Contract 1224666 issued by JPL, Caltech,
under NASA contract 1407.

\end{document}